\documentclass[twocolumn,showpacs,preprintnumbers,amsmath,amssymb]{revtex4}

\usepackage{graphicx}
\usepackage{dcolumn}
\usepackage{bm}
\usepackage{mathrsfs}
\usepackage{color}

\newcommand{\bvec}[1]{\mbox{\boldmath $#1$}}

\begin{document}

\title{Effect of cluster-shell competition of $^{12}\mathrm{C}$ 
on $E0$ transitions in  $^{16}\mathrm{O}$}

\author{H.~Matsuno$^1$ and N.~Itagaki$^2$}

\affiliation{$^1$Department of Physics, Kyoto University,\\
Kitashirakawa Oiwake-Cho, Kyoto 606-8502, Japan\\
$^2$Yukawa Institute for Theoretical Physics, Kyoto University,\\
Kitashirakawa Oiwake-Cho, Kyoto 606-8502, Japan}

\date{\today}

\begin{abstract}
In $^{16}\mathrm{O}$, we investigate the relation between 
the $E0$ monopole transition matrix elements and cluster-shell competition 
using antisymmetrized quasi cluster model (AQCM), where
the dissolution of $\alpha$ clusters into quasi clusters due to the effect of the spin-orbit force 
is introduced.
We focus on the structure change when the strength of the spin-orbit force is varied.
The ground state dominantly has a compact four $\alpha$ structure (doubly magic structure of the $p$ shell), which is rather independent of the strength.
However the first excited state ($0_2^+$) has $^{12}\mathrm{C}+\alpha$ cluster structure and
this state is largely affected by the increase of the strength;
the subclosure configuration of the $p_{3/2}$ shell for the $^{12}$C cluster part
becomes more important than three $\alpha$ configuration.
The $E0$ transition from the ground state with compact four $\alpha$ configuration to the $0^+_2$ state with $^{12}\mathrm{C}+\alpha$ cluster configuration is suppressed when increasing 
the spin-orbit strength.
Although 
the $E0$ operator itself does not have the spin dependence, the matrix element
is sensitive to the difference of intrinsic spin structures of the two states.
The $E0$ transition
 characterizes the persistence of four $\alpha$  structure in the $0^+_2$ state.
\end{abstract}


\maketitle

\section{introduction}
The $E0$ monopole transition has been regarded as a physical quantity which is related to the matter properties of heavy nuclei.
In addition, recently it has been discussed that it is closely related to
the structures of light nuclei, especially for their cluster structure \cite{Kawabata_PhysLettB646_6}.
The $E0$ monopole excitation induces the breathing mode of nuclei. However changing nuclear density requires highly excitation energy because of the saturation property of nuclear systems.
On the other hand, if the system is composed of strongly bound subsystems called clusters,
the relative interaction between cluster is weak, and it is possible to change the relative distances (nuclear sizes) without giving high excitation energies.
$\alpha$ clusters are very stable and they are good candidates.
The monopole excitation is expected to induce $\alpha$ cluster excitation (excitation from the ground state to states with $\alpha$ cluster structure) in low excitation energy region. Indeed,
strong monopole transition and the relation to $\alpha$ clustering have been discussed 
in $^{10}\mathrm{B}$ \cite{Kawabata_PhysLettB646_6}, $^{12, 13}\mathrm{C}$
\cite{Sasamoto_ModPhysLettA21_2393,Yoshida_PhysRevC79_034308,Yamada_PhysRevC92_034326}, and $^{24}\mathrm{Mg}$ \textit{etc} \cite{Ichikawa_PhysRevC83_061301,Ichikawa_PhysRevC86_031303}. \par
Now we focus on the case that the counterpart of the $\alpha$ cluster has more complex structure.
There have been numerous works showing that the low-lying states of
$^{16}\mathrm{O}$, including the $0_2^+$ state, have $^{12}\mathrm{C}+\alpha$ structure \cite{Horiuchi_ProgTheorPhys40_277,Arima_PhysLettB24_129,Haxton_PhysRevLett65_1325}.
Also in this nucleus, the $E0$ strength distribution has been observed \cite{Lui_PhysRevC64_064308}.  If we compare the results with the random phase approximation (RPA) calculation based on the mean-field theory \cite{Ma_PhysRevC55_2385}, the agreement 
in high excitation energy region is good; however it is not in low excitation energy region. 
There some of the strong peaks observed are missing in the calculation.
For example, the $E0$ transitions from the $0_1^+$ to the $0_2^+$ and $0_3^+$ states are observed to be $3.55\pm0.21\,e\,\mathrm{fm}^2$ and $4.03\pm0.09\,e\,\mathrm{fm}^2$, respectively \cite{Ajzenberg_NuclPhysA460_1,Tilley_NuclPhysA564_1}, and they correspond to about 3\% and 8\% of the energy weighted sum rule \cite{Yamada_ProgTheorPhys120_1139}.
Based on the cluster approaches, these $E0$ transitions to low-lying states are theoretically explained in the frameworks of $^{12}\mathrm{C}+\alpha$ orthogonality condition model (OCM) \cite{Suzuki_ProgTheorPhys55_1751}, $4\alpha$ OCM \cite{Yamada_PhysRevC85_034315} and 
$^{12}\mathrm{C}$ (antisymmetrized molecular dynamics: AMD) $+\alpha$ generator coordinate method (GCM) \cite{En'yo_PhysRevC89_024302}.
Yamada \textit{et al}. pointed out in Ref.~\cite{Yamada_ProgTheorPhys120_1139} that there are two reasons for the $E0$ transition from the ground state to low-lying cluster states.
One is the duality; simple shell model wave function automatically contains the clustering degree of freedom.
If we take the zero limit for the relative distance between $^{12}\mathrm{C}$ and $\alpha$, the wave function agrees with the closed $p$ shell configuration of the shell model. 
Thus there is certain path from the ground state to the  $^{12}\mathrm{C}+\alpha$ cluster configuration.
This is explained by the so-called Bayman-Bohr theorem \cite{Bayman_NuclPhys9_596}.
The other is ground state correlation.
The ground state slightly deviates from the shell model limit and contains more $^{12}\mathrm{C}+\alpha$ cluster configurations when the wave functions corresponding to this path are superposed. 
This effect enhances the transition from the ground
state to the $^{12}\mathrm{C}+\alpha$ cluster states. \par
However, the discussion of Bayman-Bohr theorem is based on the three-dimensional harmonic oscillator type wave function (or $N\alpha$ cluster model wave function with some limit of relative distances) and it is not trivial whether the same logic holds or not when the $jj$-coupling shell model wave functions are introduced. The ground state is rather safe;
the closed $p$ shell configuration, which plays a dominant role in the ground state, can be equally described by the  four $\alpha$ cluster model and $jj$-coupling shell model.
On the contrary, there appears non-negligible difference in the wave function of
$^{12}\mathrm{C}+\alpha$ states when the $jj$-coupling $^{12}$C is introduced.
In our previous analysis for $^{12}\mathrm{C}$, we discussed that the ground state of $^{12}\mathrm{C}$ 
is an intermediate state between three $\alpha$ cluster state and the subclosure configuration of $p_{3/2}$ in the $jj$-coupling shell model, and there the contribution of the spin-orbit force is quite strong \cite{Suhara_PhysRevC87_054334}.
In $^{16}$O, both wave functions of the ground and excited states contributes to the monopole transition matrix elements, and it is worthwhile to investigate whether the inclusion of shell model like wave functions for the $^{12}$C cluster part changes the story of monopole transition strengths in the low energy regions or not.\par
In this paper, we discuss  the behavior of the $E0$ transition
matrix elements from the ground state when the $jj$-coupling shell model 
$^{12}$C is mixed with three $\alpha$ 
in the excited states of  $^{16}\mathrm{O}$.
We take notice on the $E0$ transition matrix elements as a function of the strength of the spin-orbit force
in the Hamiltonian. With increasing the spin-orbit strength, the $\alpha$ breaking components 
become more important, and $jj$-coupling shell model $^{12}$C strongly
mixes in the $^{12}$C+$\alpha$ cluster states. 
Although the $E0$ operator itself does not have the spin dependence,
the $E0$ transition matrix elements  turns out to be sensitive to this change.\par
As a theoretical framework,
we use antisymmetrized quasi cluster model (AQCM) \cite{Itagaki_PhysRevC71_064307,Suhara_PhysRevC87_054334,Masui_PhysRevC75_054309,Itagaki_PhysRevC83_014302}.
AQCM is the method, which can describe $jj$-coupling shell model wave function by extending cluster model.
The remarkable advantage of AQCM is that the number of the parameters required to 
characterize the transition from cluster state to $jj$-coupling shell state is quite small.
In addition to the distance parameters used in traditional cluster models,
AQCM needs only one new parameter $\varLambda$, which describes the dissolution of $\alpha$ clusters and change into $jj$-coupling shell state; the clusters with
finite $\varLambda$ value are called ``quasi clusters''.
In $^{12}\mathrm{C}$,
Suhara \textit{et al}. were successful to describe
the change of three $\alpha$ cluster state into the $p_{3/2}$ subclosure configuration 
of the $jj$-coupling shell model 
with only two parameters, namely distance parameter $R$ and dissolution parameter 
$\varLambda$ \cite{Suhara_PhysRevC87_054334}.
In this work, we adopt the method to describe the $^{12}\mathrm{C}$ 
part of the $^{12}\mathrm{C}+\alpha$ cluster states in $^{16}\mathrm{O}$.\par
This paper is organized as follows.
We explain our formulation for this work in Sec. \ref{formulation}.
The results and discussion are given in Sec. \ref{resultanddiscussion}.
Finally, we present conclusion and outlook in Sec. \ref{conclusion}.

\section{formulation}
\label{formulation}
In this section, we explain the wave function and Hamiltonian in our model.
\subsection{Wave function}
\subsubsection{Single-particle wave function}
The single-particle wave function is described by Gaussian packet,
\begin{align}
\phi_i=\left(\frac{2\nu}{\pi}\right)^{\frac{3}{4}}\exp[-\nu(\bvec{r}-\bvec{\zeta}_i)^2]\chi_i\tau_i,
\end{align}
where $\chi_i$ and $\tau_i$ are spin and isospin part of the $i$th single-particle wave function, respectively.
For the width parameter $\nu$  $ (=1/2b^2)$, we use the value of $b=1.6\,\mathrm{fm}$ to reproduce the radius of $^{16}\mathrm{O}$.
If we choose the same Gaussian center parameter $\zeta_i$ for four nucleons (spin up proton, spin down proton, spin up neutron and spin down neutron), the four nucleons are regarded as forming an $\alpha$ cluster.\par
\begin{figure}[tb]
\includegraphics[width=.3\textwidth]{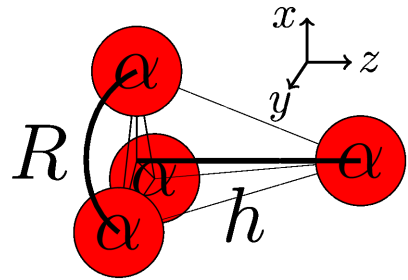}
\caption{(Color online) The schematic figure for the definitions of $R$ and $h$. 
The red spheres show the $\alpha$ clusters and three of them on the $x$-$y$ plane are changed into quasi clusters.}
\label{Fig.R.and.h}
\end{figure}
The coordinate system is defined in the following way.
Firstly, we place three $\alpha$ clusters
on the $x$-$y$ plane
in regular triangle shape.
which are regarded as $^{12}\mathrm{C}$ cluster.
The length of one side is defined as $R$.
Secondly, we place the fourth $\alpha$ cluster on the $z$ axis.
The length between the center of masses of the fourth $\alpha$ cluster and $^{12}$C is defined as $h$.
If $h=\sqrt{2/3}\times{}R$, the four $\alpha$ clusters configure tetrahedron shape.
The definitions of $R$ and $h$ are schematically shown in Fig.~\ref{Fig.R.and.h}.
Thirdly, we introduce a dissolution parameter $\varLambda$ for the three $\alpha$ clusters on the $x$-$y$ plane.
For the details of introducing $\varLambda$ for $^{12}$C, see Ref.~\cite{Suhara_PhysRevC87_054334}.
Finally, the center of gravity of the whole system is moved to the origin.

\subsubsection{Wave function of the total system}
The wave function of the total system is parity and angular momentum eigenstate:
\begin{align}
\varPhi=&\sum_{i,j,k}c_{ijk}\varPsi_{ijk},
\label{varPhi} \\
\varPsi_{ijk}=&\varPsi(R_i,h_j,\varLambda_k) \nonumber \\
=&\hat{P}^J_{MK}\hat{P}^\pi\mathcal{A}[\phi_1\cdots\phi_{16}],
\label{varPsi_ijk}
\end{align}
where $\hat{P}^J_{MK}$ and $\hat{P}^\pi$ are angular momentum projection operator and parity projection operator, respectively.
$\mathcal{A}$ is antisymmetrizer for all sixteen nucleons.
The parameters are taken as $\{R_i\}=0.1, 1.0, 2.0, 3.0, 4.0\,\mathrm{fm}$, $\{h_j\}=\sqrt{2/3}\times(0.1, 1.0, 2.0, 3.0, 4.0, 5.0, 6.0, 7.0, 8.0)\,\mathrm{fm}$ and $\{\varLambda_k\}=0, 1/3, 2/3, 1$, respectively.
There are $5\times9\times4=180$ bases and the coefficient $c_{ijk}$ is determined by solving the Hill-Wheeler equation.\par
\begin{figure}[tb]
\includegraphics[width=.42\textwidth]{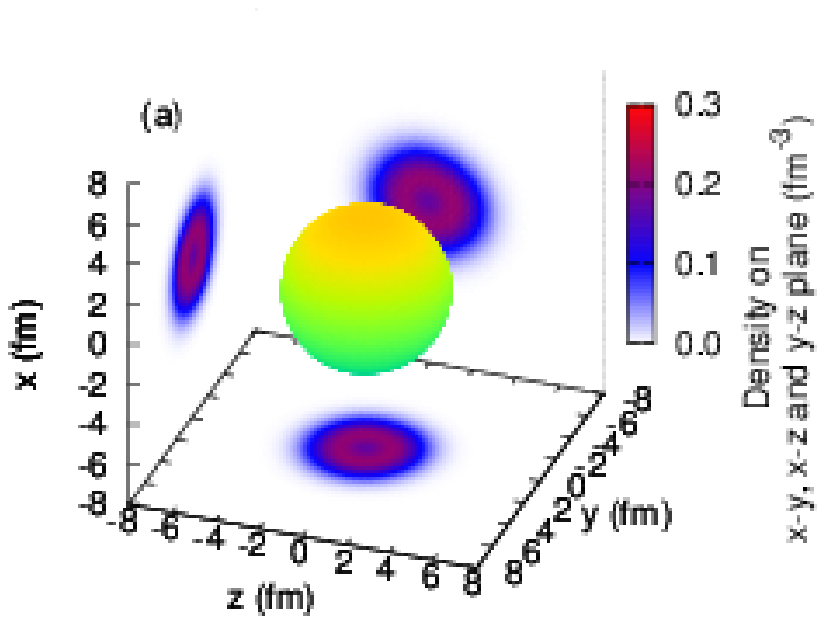}
\includegraphics[width=.42\textwidth]{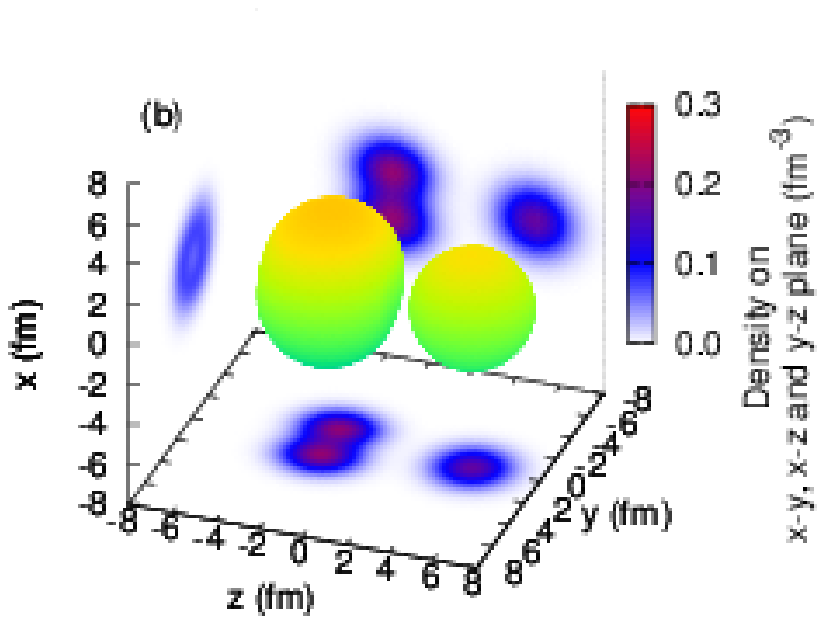}
\includegraphics[width=.42\textwidth]{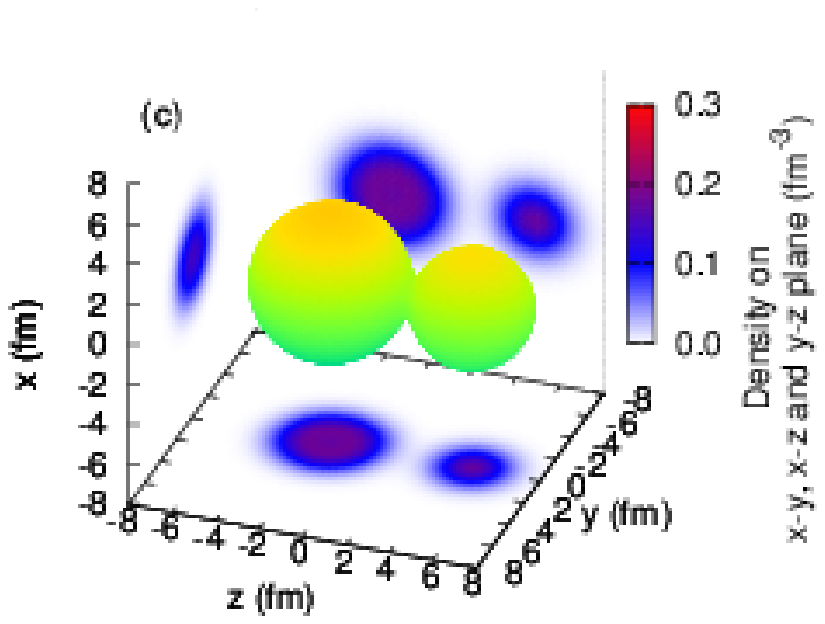}
\caption{(Color online) Intrinsic density distributions of typical basis states.
The nuclear surface is defined as the point with the density of $0.01\,\mathrm{fm}^{-3}$.
Right back, left back and bottom figures are density distributions on the $x$-$z$ plane, $x$-$y$ plane and $y$-$z$ plane, respectively.
These parameters are (a) $R=0.1\,\mathrm{fm}$,  $h=\sqrt{2/3}\times0.1\,\mathrm{fm}$, $\varLambda=0$ , (b) $R=0.1\,\mathrm{fm}$, $h=\sqrt{2/3}\times8.0\,\mathrm{fm}$, $\varLambda=0$ and (c) $R=0.1\,\mathrm{fm}$, $h=\sqrt{2/3}\times8.0\,\mathrm{fm}$, $\varLambda=1$.}
\label{Fig.Density}
\end{figure}
The intrinsic density distributions (\textit{i.e.} the density distributions before angular momentum and parity projections) of typical basis states are shown in Fig.~\ref{Fig.Density}.
In Fig.~\ref{Fig.Density} (a), the four $\alpha$ clusters configure compact tetrahedron shape,
and parameters are set to $R=0.1\,\mathrm{fm}$,  $h=\sqrt{2/3}\times0.1\,\mathrm{fm}$ and $\varLambda=0$.
In this case, the density looks having a spherical symmetry.
This state corresponds to the $(0s)^4(0p)^{12}$ closed shell configuration at the limit of $h=\sqrt{2/3}\times{}R\to0$ as described in the so-called Bayman-Bohr theorem \cite{Bayman_NuclPhys9_596}; the zero limit
for the distances between $\alpha$ clusters with certain shape corresponds to the closed-shell configurations.
Figure~\ref{Fig.Density} (b) and (c) 
show the $^{12}\mathrm{C}+\alpha$ cluster like states, where
the parameter $h$ is increased to $h=\sqrt{2/3}\times8.0\,\mathrm{fm}$, 
while keeping $R=0.1\,\mathrm{fm}$.
In Fig.~\ref{Fig.Density} (b) and (c), $^{12}\mathrm{C}$ cluster is put on the left-hand side and the 
last $\alpha$ cluster is on the right-hand side.
In Fig.~\ref{Fig.Density} (b), the dissolution parameter is set to $\varLambda=0$ and $^{12}$C 
cluster is nothing but three $\alpha$ clusters.
In Fig.~\ref{Fig.Density} (c), the dissolution parameter is set to $\varLambda=1$ and three $\alpha$ clusters 
in $^{12}$C are
changed into quasi clusters, which correspond to
the subclosure configuration of $p_{3/2}$ at the limit of $R \to 0$.
Here, the density of the $^{12}\mathrm{C}$ cluster part looks having a spherical symmetry.

\subsection{Hamiltonian}
The Hamiltonian used in our work is given as
\begin{align}
\hat{H}=\hat{T}-\hat{T}_\mathrm{G}+\hat{V}_\mathrm{C}+\hat{V}_\mathrm{LS}+\hat{V}_\mathrm{Coulomb},
\end{align}
where $\hat{T}$ is total kinetic energy operator and $\hat{T}_\mathrm{G}$ is kinetic energy of center of mass motion.
We use the Volkov No.2 \cite{Volkov_NuclPhys74_33} for the central force $\hat{V}_\mathrm{C}$:
\begin{align}
\hat{V}_\mathrm{C}=&\sum_{i<j}^A\left[V_a\exp\left(-\frac{\hat{\bvec{r}}_{ij}^2}{\alpha^2}\right)+V_r\exp\left(-\frac{\hat{\bvec{r}}_{ij}^2}{\rho^2}\right)\right] \nonumber \\
&\times\left[W+B\hat{P}^\sigma_{ij}-H\hat{P}^\tau_{ij}-M\hat{P}^\sigma_{ij}\hat{P}^\tau_{ij}\right],
\end{align}
where $V_a=-60.65\,\mathrm{MeV}$, $V_r=61.14\,\mathrm{MeV}$, $\alpha=1.80\,\mathrm{fm}$ and $\rho=1.01\,\mathrm{fm}$ are original values and we adopt $M=1-W=0.62$ and $B=H=0.125$.
We use spin-orbit part of G3RS force \cite{Tamagaki_ProgTheorPhys39_91} for $\hat{V}_\mathrm{LS}$:
\begin{align}
\hat{V}_\mathrm{LS}=&\sum_{i<j}^A\left[V_\mathrm{LS1}\exp\left(-\frac{\hat{\bvec{r}}_{ij}^2}{\eta_1^2}\right)+V_\mathrm{LS2}\exp\left(-\frac{\hat{\bvec{r}}_{ij}^2}{\eta_2^2}\right)\right] \nonumber \\
&\times\hat{P}_{ij}(^3O)\hat{\bvec{L}}_{ij}\cdot\hat{\bvec{S}}_{ij},
\end{align}
where $\eta_1=0.447\,\mathrm{fm}$ and $\eta_2=0.6\,\mathrm{fm}$ are original values and $V_\mathrm{LS}\equiv{}V_\mathrm{LS1}=-V_\mathrm{LS2}$ is variable parameter in our work.
$\hat{V}_\mathrm{Coulomb}$ is Coulomb potential for the protons.

\section{result and discussion}
\label{resultanddiscussion}
\subsection{Effects of spin-orbit strength}
In this section, we discuss the $V_\mathrm{LS}$ (strength of the spin-orbit force) dependence of the $0^+$ energy levels and $E0$ transition matrix elements.
The reasonable $V_\mathrm{LS}$ value of $1600\ \sim\ 2000\,\mathrm{MeV}$ has been suggested 
in the study of scattering phase shift of $\alpha+n$ system \cite{Okabe_ProgTheorPhys61_1049}.
However here
we set up two extreme cases of $V_\mathrm{LS} =$ 0 MeV and 3000 MeV and vary  $V_\mathrm{LS}$
between them to see the tendency of the energy levels and $E0$ transition matrix elements.
We discuss the change of the wave functions
by calculating squared overlap between the final solution and each basis state.

\subsubsection{Energy levels}
\begin{figure}[tb]
\centering
\includegraphics[width=.48\textwidth]{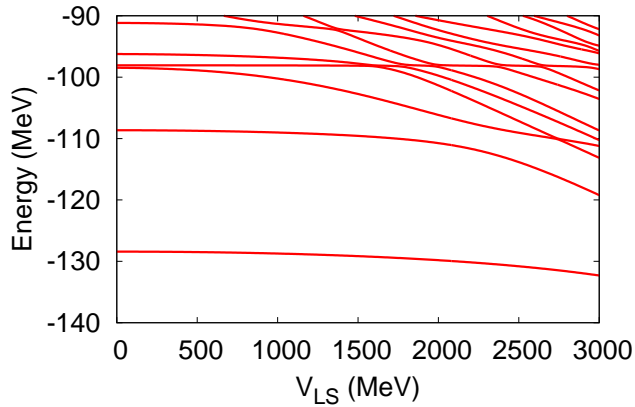}
\caption{(Color online) Energy levels of $0^+$ states in $^{16}$O 
against the strength of the spin-orbit force, 
$V_\mathrm{LS}$.}
\label{Fig.EnergyLS}
\end{figure}
Firstly, we investigate the $0^+$ energy levels of $^{16}$O.
The $0^+$ energy levels as a function of the strength of the spin-orbit force, $V_\mathrm{LS}$, 
are shown in Fig.~\ref{Fig.EnergyLS}.
The experimental ground state energy is $-127.6\,\mathrm{MeV}$ \cite{NNDC}.
Here we plot points corresponding to the solution of the Hill-Wheeler equation for each fixed $V_\mathrm{LS}$ value and connect them with lines in order from bottom.
The ``$n$th state'' is defined at each $V_\mathrm{LS}$ value from the bottom.
For $0\le{}V_\mathrm{LS}\le1500\,\mathrm{MeV}$, the energies of calculated first and second states are almost independent of $V_\mathrm{LS}$.
For $V_\mathrm{LS}\ge2000\,\mathrm{MeV}$, the energy of calculated first state slightly decreases with increasing $V_\mathrm{LS}$ value; however the second state is much more influenced.
This is because the dominant configuration of the ground state is the closed $p$ shell configuration, where the contribution of the spin-orbit force to $p_{3/2}$ and $p_{1/2}$ cancels.
On the other hand, in calculated second state, the spin-orbit force acts attractively for the $^{12}$C cluster part. 
In the region slightly above $V_\mathrm{LS} = 2000$ MeV, level repulsion of calculated second and third states occurs.\par
Experimentally the $0_2^+$ state is observed at $E_x = 6.05\,\mathrm{MeV}$ \cite{Ajzenberg_NuclPhysA460_1,Tilley_NuclPhysA564_1}. However, this excitation energy is 
calculated to be higher by more than $10\,\mathrm{MeV}$ without the spin-orbit force at $V_\mathrm{LS} = 0\,\mathrm{MeV}$.
This has been a long standing problem of traditional (microscopic) $\alpha$ cluster models, which cannot take into account the spin-orbit effect. Now this is considerably improved by introducing the dissolution of $\alpha$ clusters.\par
As discussed later, the main component of the calculated second state at $V_\mathrm{LS}=0\,\mathrm{MeV}$ is four $\alpha$ clusters; even after allowing the dissolution of $\alpha$ clusters, 
basis states 
with finite $\varLambda$ values
do not contribute to the calculated second state without spin-orbit force, and $^{12}$C cluster part is nothing but three $\alpha$ clusters.
It is intriguing to point out that even in the region of $V_\mathrm{LS}\ge2000\,\mathrm{MeV}$, such four
 $\alpha$ like state survives and the energy stays almost constant.
Finally the four $\alpha$ state becomes calculated fourth state at $V_\mathrm{LS}=3000\,\mathrm{MeV}$.
The result suggests that the four $\alpha$ state is rather decoupled from other states with finite $\varLambda$ values and survives even after switching on the spin-orbit force.\par
We also find another example of such decoupling of four $\alpha$ state at $-98.1$ MeV.
At $V_\mathrm{LS}=0$, the calculated state is fourth state.
The energy of this state does not change even after increasing $V_\mathrm{LS}$.
Many level crossings with other states occur; however the character of this state
remains at this energy. 

\subsubsection{$E0$ transition matrix element}
Next we show the $E0$ transition from the ground state and discuss the effect of $\alpha$ dissolution due to the spin-orbit force. 
The $E0$ transition matrix element $M(E0, 0_n^+-0_1^+)$ from the $0_1^+$ to the $0_n^+$ state is defined as
\begin{align}
M(E0, 0_n^+-0_1^+)\equiv\left|\left\langle{}0_n^+\left|\sum_{i=1}^{16}e\frac{1+\hat{\tau}_{i3}}{2}(\hat{\bvec{r}}_i-\hat{\bvec{r}}_\mathrm{c.m.})^2\right|0_1^+\right\rangle\right|,
\end{align}
where $\hat{\bvec{r}}_\mathrm{c.m.}\equiv\frac{1}{16}\sum_{i=1}^{16}\hat{\bvec{r}}_i$ is the center-of-mass coordinate operator.\par
\begin{figure}[tb]
\centering
\includegraphics[width=.48\textwidth]{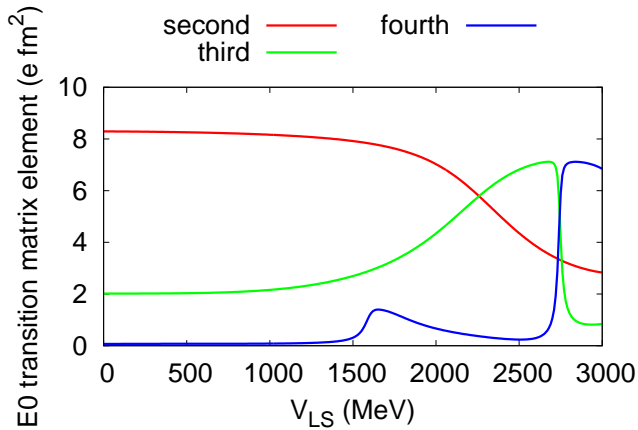}
\caption{(Color online) $E0$ transition matrix elements from the first 
$0^+$ state of $^{16}$O 
to the second, third and fourth states calculated in Fig.~\ref{Fig.EnergyLS}.
The horizontal axis is the strength of the spin-orbit force, $V_\mathrm{LS}$.}
\label{Fig.MonopoleLS}
\end{figure}
The  $E0$ transition matrix elements from the first 
to the second, third and fourth $0^+$ states calculated by changing
the $V_\mathrm{LS}$ value,
are shown in Fig.~\ref{Fig.MonopoleLS}.
Here, $n$th state is defined  from the bottom at each $V_\mathrm{LS}$ value. 
For $0\le{}V_\mathrm{LS}\le1500\,\mathrm{MeV}$, the $E0$ transition matrix element from the first state to the second state is around 8 $e$~fm$^2$ almost independent of $V_\mathrm{LS}$.
This value is much larger than the observed one ($3.55\pm0.21\,e~\mathrm{fm}^2$).
In the region of
$V_\mathrm{LS}\ge1500\,\mathrm{MeV}$, the $E0$ transition matrix element from the first state to the second state decreases. 
On the other hand, the one from the first $0^+$ state to the third $0^+$ state increases (the observed value for $0^+_3$ is $4.03\pm0.09\,e~\mathrm{fm}^2$).
This is due to the interchanges of the wave functions of the second  and third states after their level repulsion. 
Why these changes influence the $E0$ transition probability will be discussed later in detail.
Experimentally the transition matrix element to the third $0^+$ state is slightly larger
than one for the second $0^+$ state,
and this is realized in our model with the $V_\mathrm{LS}$ value slightly above 2000 MeV.

\subsubsection{Squared overlap}
In this subsection, the wave function of each state is analyzed in more detail.
The squared overlap between the final solution, $|\varPhi\rangle$ in Fig.~\ref{Fig.EnergyLS}, and each basis state, $|\varPsi_{ijk}\rangle$ in Eqs. (\ref{varPhi}) and (\ref{varPsi_ijk}), at fixed $V_\mathrm{LS}$ value is defined as
\begin{align}
\left|\frac{\langle\varPsi_{ijk}|\varPhi\rangle}{\sqrt{\langle\varPsi_{ijk}|\varPsi_{ijk}\rangle}\sqrt{\langle\varPhi|\varPhi\rangle}}\right|^2,
\end{align}
and we compare the squared overlaps for the cases of $V_\mathrm{LS}=0\,\mathrm{MeV}$ and $V_\mathrm{LS}=3000\,\mathrm{MeV}$.\par

\begin{figure}[tb]
\centering
\includegraphics[width=.40\textwidth]{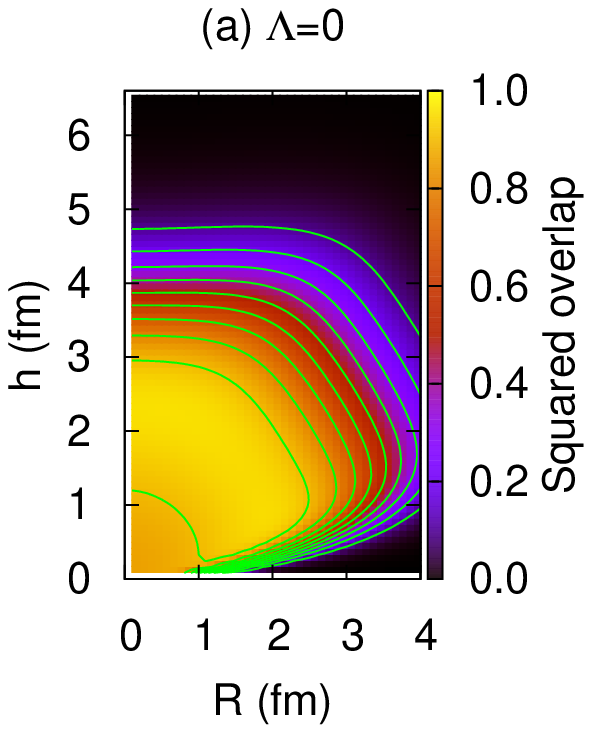}
\includegraphics[width=.40\textwidth]{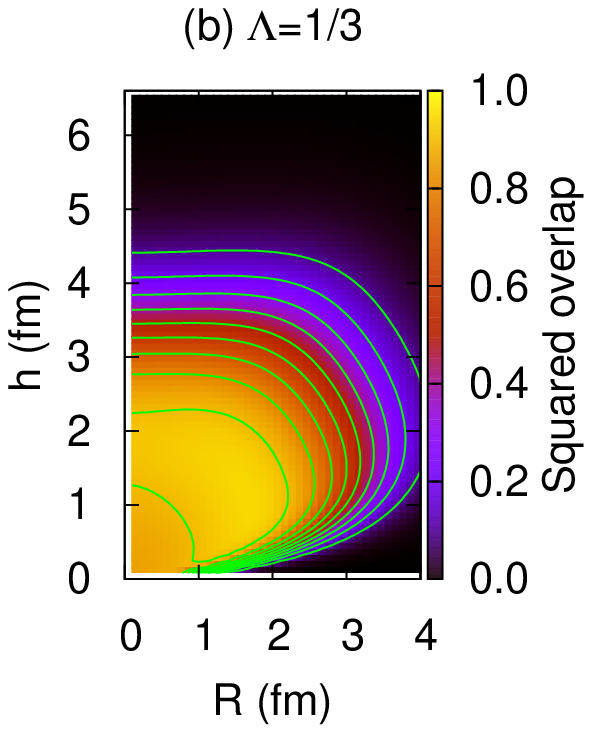}
\includegraphics[width=.40\textwidth]{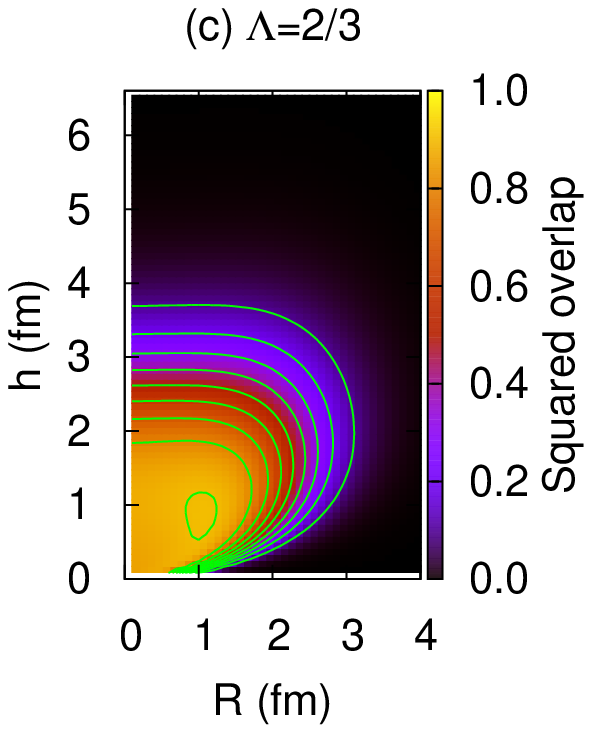}
\includegraphics[width=.40\textwidth]{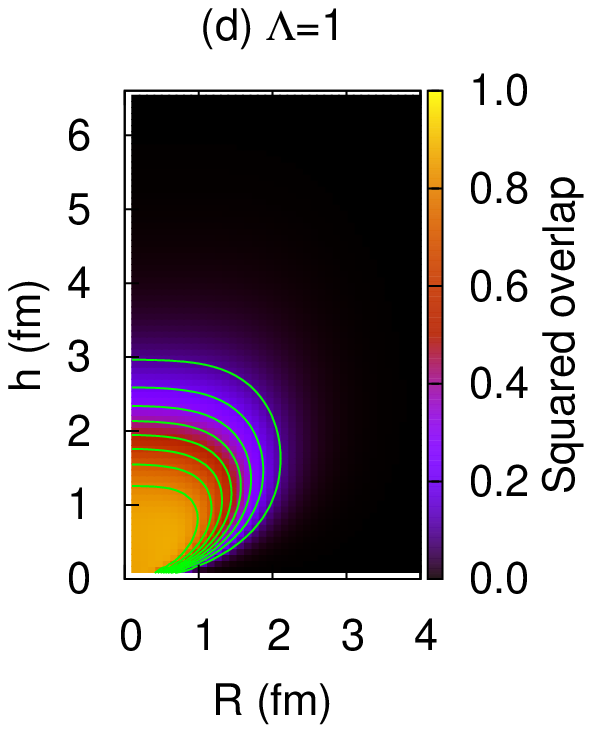}
\caption{(Color online) Squared overlap between the first state 
calculated at $V_\mathrm{LS}=0\,\mathrm{MeV}$
in Fig.~\ref{Fig.EnergyLS} and the basis state $|\varPsi(R,h,\varLambda)\rangle$.}
\label{Fig.OverlapLS01st}
\end{figure}
\begin{figure}[tb]
\centering
\includegraphics[width=.40\textwidth]{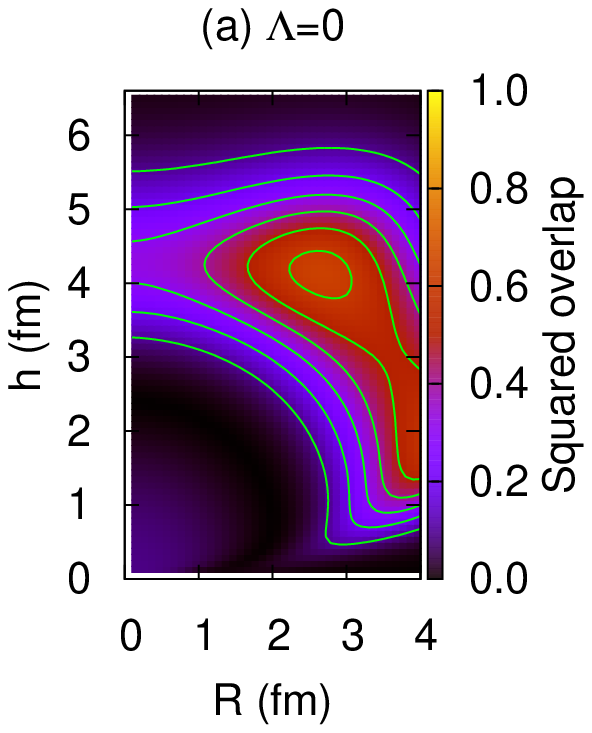}
\includegraphics[width=.40\textwidth]{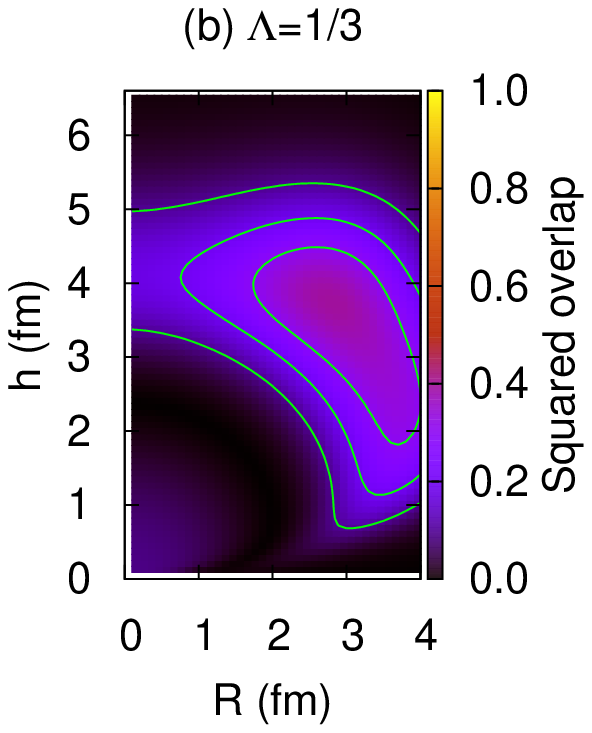}
\includegraphics[width=.40\textwidth]{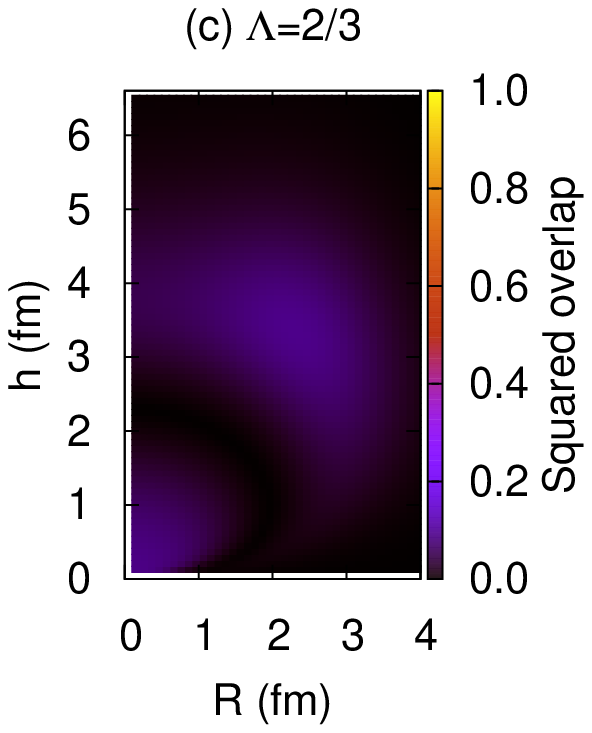}
\includegraphics[width=.40\textwidth]{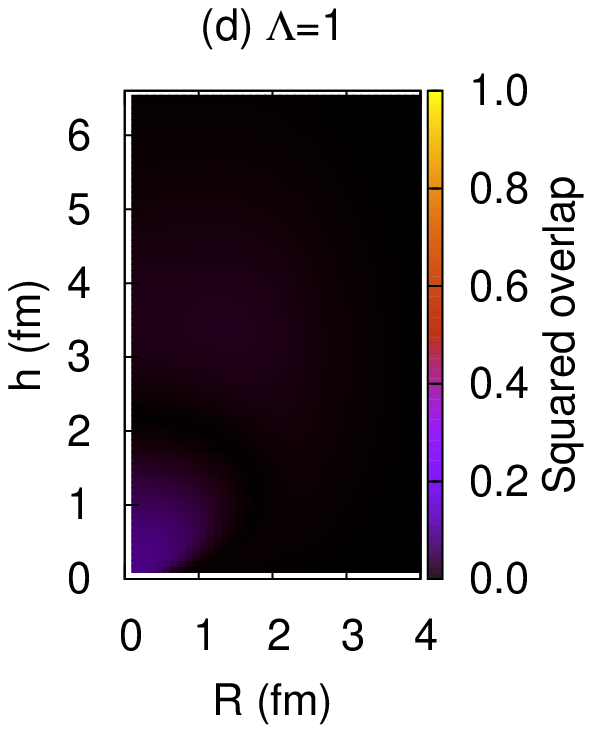}
\caption{(Color online) Squared overlap between the second state 
calculated at $V_\mathrm{LS}=0\,\mathrm{MeV}$
in Fig.~\ref{Fig.EnergyLS} and the basis state $|\varPsi(R,h,\varLambda)\rangle$.}
\label{Fig.OverlapLS02nd}
\end{figure}
\begin{figure}[tb]
\centering
\includegraphics[width=.40\textwidth]{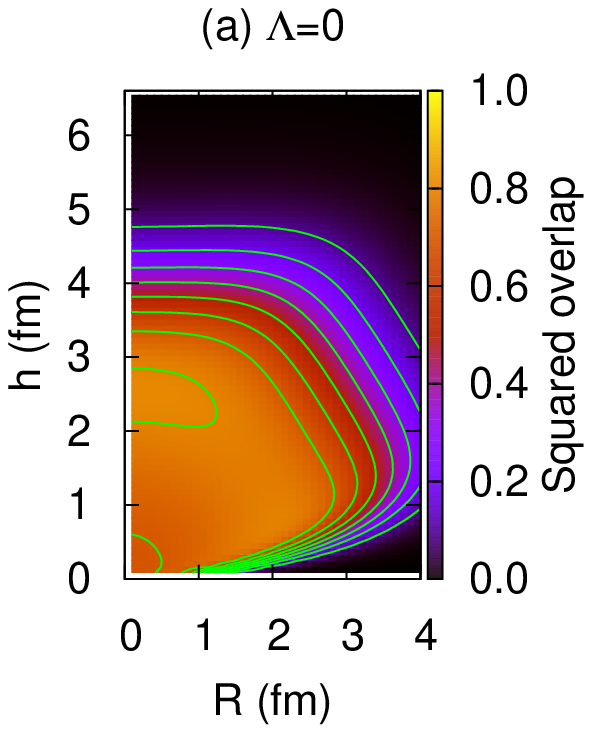}
\includegraphics[width=.40\textwidth]{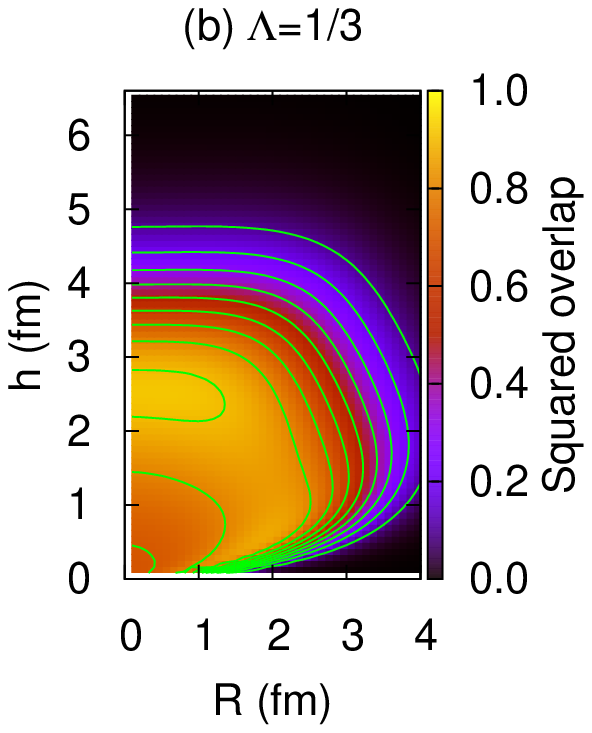}
\includegraphics[width=.40\textwidth]{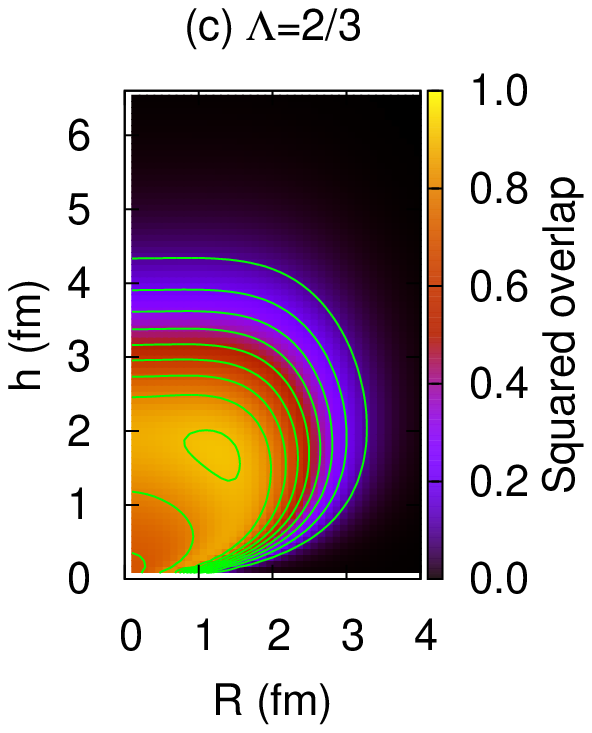}
\includegraphics[width=.40\textwidth]{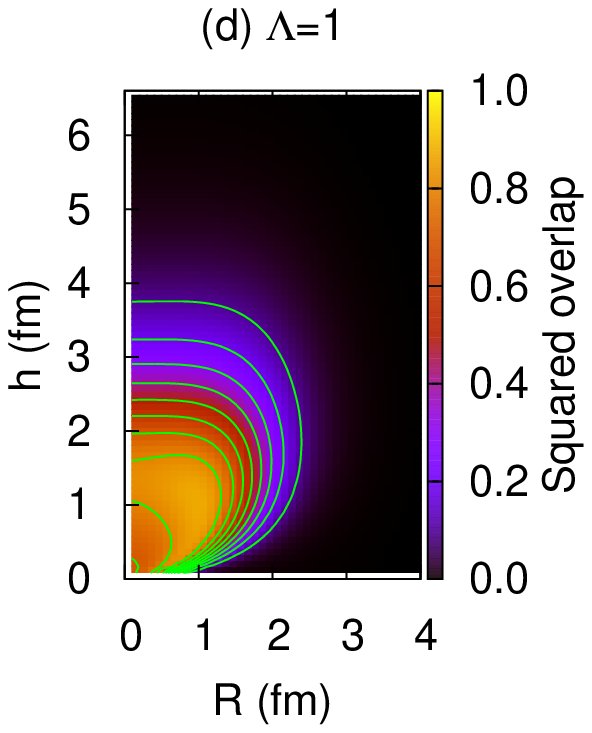}
\caption{(Color online) Squared overlap between the first state calculated 
at $V_\mathrm{LS}=3000\,\mathrm{MeV}$
in Fig.~\ref{Fig.EnergyLS} and the basis state $|\varPsi(R,h,\varLambda)\rangle$.}
\label{Fig.OverlapLS30001st}
\end{figure}
\begin{figure}[tb]
\centering
\includegraphics[width=.40\textwidth]{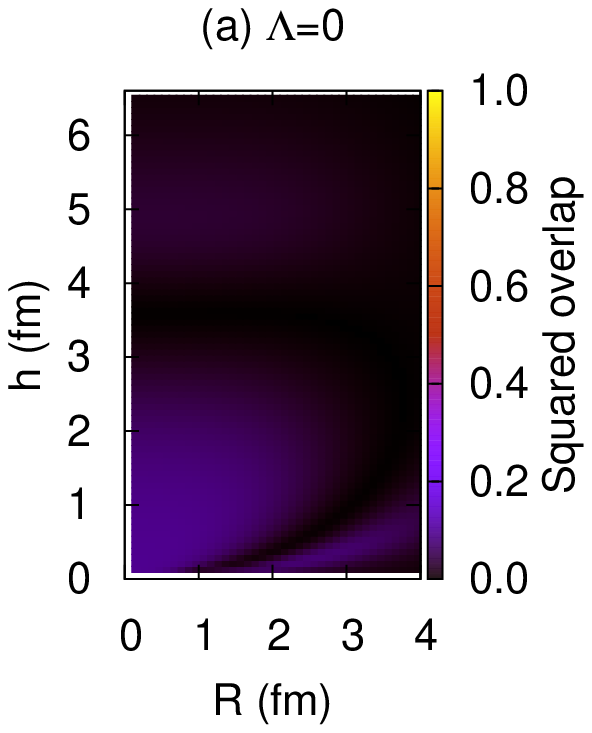}
\includegraphics[width=.40\textwidth]{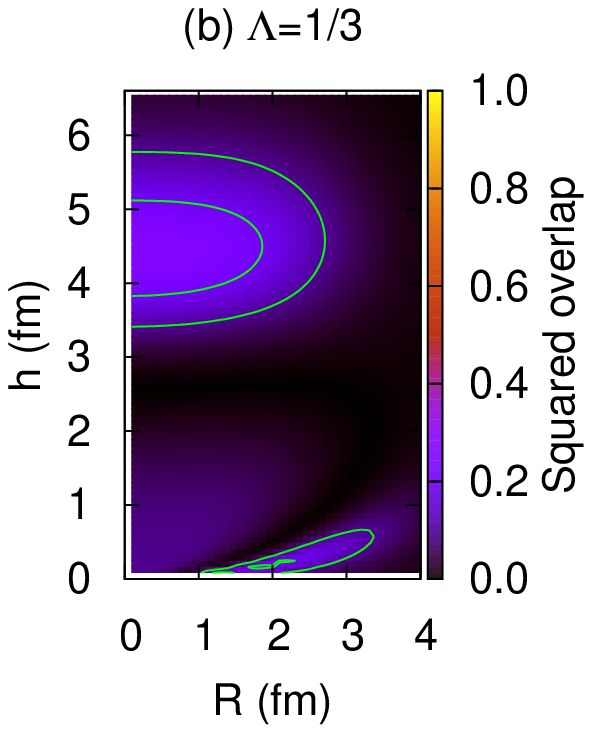}
\includegraphics[width=.40\textwidth]{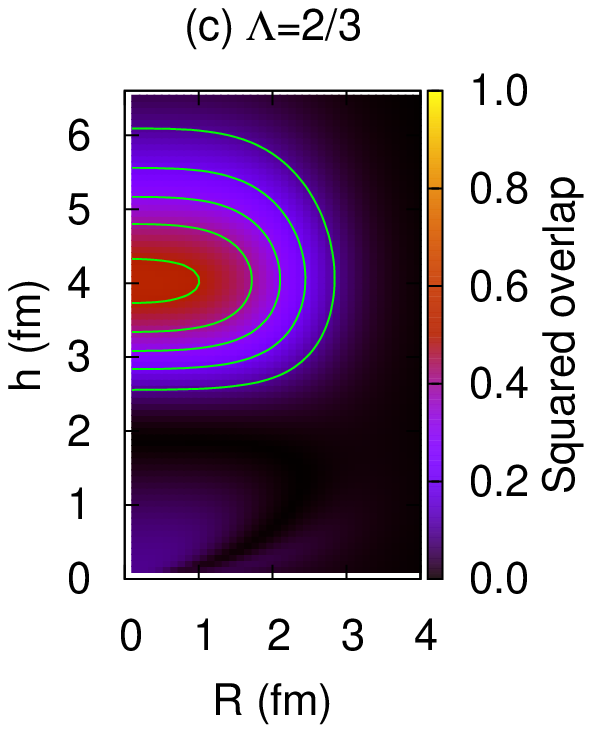}
\includegraphics[width=.40\textwidth]{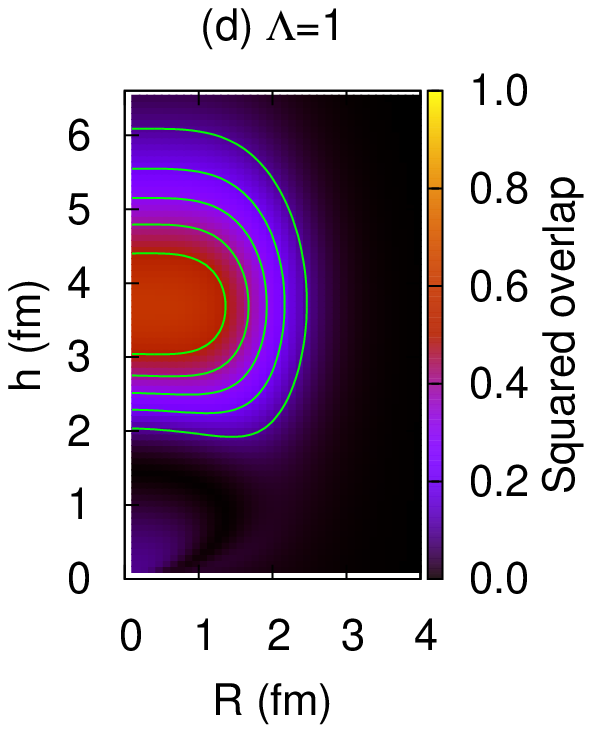}
\caption{(Color online) Squared overlap between the second state calculated 
at $V_\mathrm{LS}=3000\,\mathrm{MeV}$
in Fig.~\ref{Fig.EnergyLS} and the basis state $|\varPsi(R,h,\varLambda)\rangle$.}
\label{Fig.OverlapLS30002nd}
\end{figure}
\par 
We start with the case without the spin-orbit force ($V_\mathrm{LS}=0\,\mathrm{MeV}$).
The squared overlap between the first state calculated 
at $V_\mathrm{LS}=0\,\mathrm{MeV}$
in Fig.~\ref{Fig.EnergyLS} and $|\varPsi_{ijk}\rangle$  is shown in Fig.~\ref{Fig.OverlapLS01st}.
Figures~\ref{Fig.OverlapLS01st} (a), (b), (c) and (d) are the squared overlaps with $\varLambda=0$, $1/3$, $2/3$ and $1$ basis states, respectively.
The first state is obtained to be a compact tetrahedral structure (small $h$ and $R$), and $\varLambda=0$ gives the largest squared overlap.\par
The squared overlap between the second state calculated 
at $V_\mathrm{LS}=0\,\mathrm{MeV}$
in Fig.~\ref{Fig.EnergyLS} and $|\varPsi_{ijk}\rangle$ is shown in Fig.~\ref{Fig.OverlapLS02nd}.
Figures~\ref{Fig.OverlapLS02nd} (a), (b), (c) and (d) are the squared overlaps with $\varLambda=0$, $1/3$, $2/3$ and $1$ basis states, respectively.
In this case, spatially expanded triangular pyramid structures (large $R$ and $h$) with $\varLambda=0$ are the dominant configurations.
Without the spin-orbit force at $V_\mathrm{LS}=0$ MeV, both the first and second states are obtained to have four $\alpha$ configurations, although their spatial extensions are quite different.\par
Next we discuss the cases calculated with extreme spin-orbit strength
 ($V_\mathrm{LS}=3000\,\mathrm{MeV}$).
The squared overlap between the first state calculated 
at $V_\mathrm{LS}=3000\,\mathrm{MeV}$ 
in Fig.~\ref{Fig.EnergyLS} and $|\varPsi_{ijk}\rangle$ 
is shown in Fig.~\ref{Fig.OverlapLS30001st}.
Figures~\ref{Fig.OverlapLS30001st} (a), (b), (c) and (d) are the squared overlaps with $\varLambda=0$, $1/3$, $2/3$ and $1$ basis states, respectively.
Comparing with the case of $V_\mathrm{LS}=0\,\mathrm{MeV}$, still compact tetrahedral structures (small $R$ and $h$) are important, but the squared overlap with $\varLambda>0$ basis states are much increased as seen in Figs.~\ref{Fig.OverlapLS30001st} (b), (c), and (d).\par
Finally, the squared overlap between the second state calculated 
at $V_\mathrm{LS}=3000\,\mathrm{MeV}$ 
in Fig.~\ref{Fig.EnergyLS} and $|\varPsi_{ijk}\rangle$ 
is shown in Fig.~\ref{Fig.OverlapLS30002nd}.
Figures~\ref{Fig.OverlapLS30002nd} (a), (b), (c) and (d) are the squared overlaps with $\varLambda=0$, $1/3$, $2/3$ and $1$ basis states, respectively.
In this case, the optimal $h$ value is rather large, but $R$ is small and basis states with $\varLambda$=1 play a dominant role. Thus the subclosure configuration of $p_{3/2}$ is realized in $^{12}\mathrm{C}$.
The second state calculated at $V_\mathrm{LS}=0\,\mathrm{MeV}$ and that at $V_\mathrm{LS}=3000\,\mathrm{MeV}$ are significantly different,
and this difference affects the $E0$ transition as discussed in detail from now.

\subsection{Mechanism for the $E0$ transition suppression}
In this subsection, we discuss how the level repulsion and interchange of the
wave functions between the second and third states calculated in Fig.~\ref{Fig.EnergyLS}  
affect the $E0$ transition probability from the first state.
We focus on the change of the wave function of the second state, 
especially for the change of $^{12}\mathrm{C}$ cluster part from three $\alpha$ clusters to the $p_{3/2}$ subclosure of the $jj$-coupling shell model.
\par
\begin{figure}[tb]
\centering
\includegraphics[width=.48\textwidth]{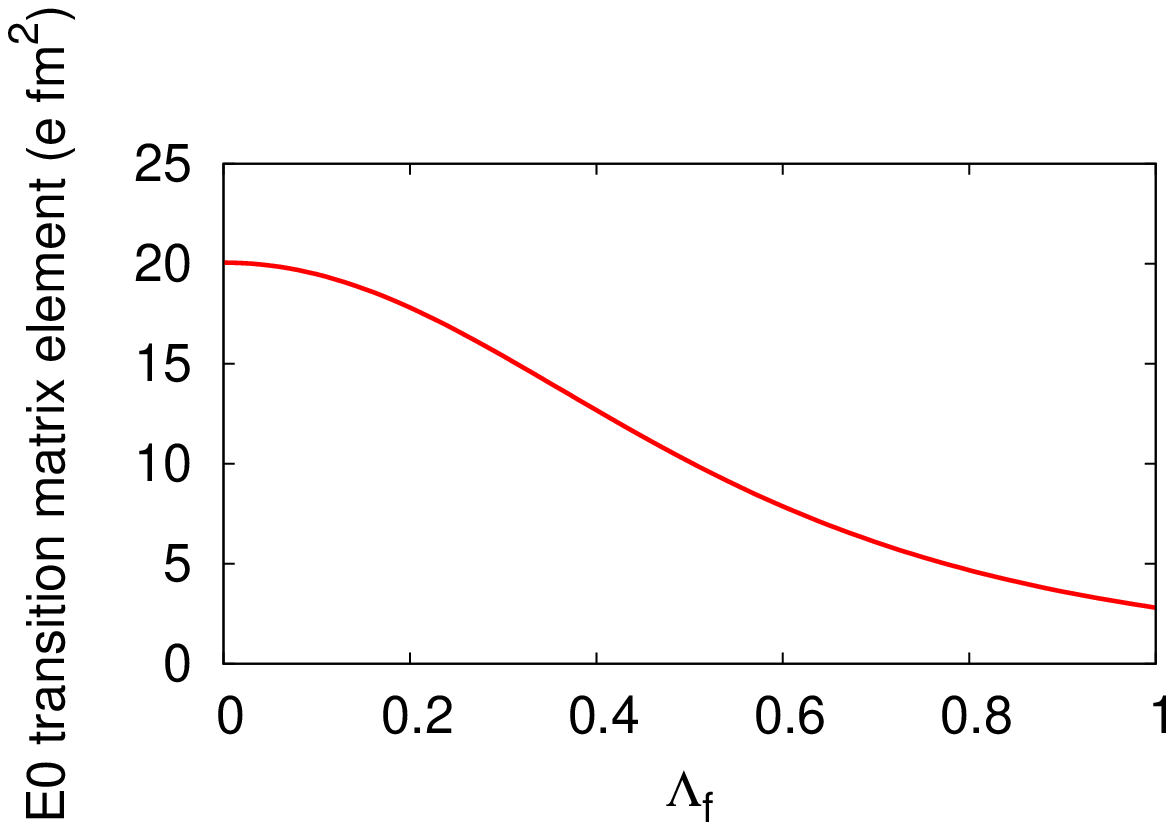}
\caption{(Color online) $E0$ transition matrix element between the compact four $\alpha$ state with $\varPsi(R=0.1\,\mathrm{fm}, h=\sqrt{2/3}\times{}0.1\,\mathrm{fm}, \varLambda=0)$ and typical $^{12}\mathrm{C}+\alpha$ cluster state $\varPsi(R=0.1\,\mathrm{fm}, h=\sqrt{2/3}\times{}5.0\,\mathrm{fm}, \varLambda=\varLambda_f)$ as a function of $\varLambda_f$.}
\label{Fig.MonoVSLambda}
\end{figure}

\subsubsection{$E0$ transition between two typical basis states}
To discuss the change of the $E0$ transition probability,
firstly we prepare typical basis states representing the characters of
the first and second states obtained in Fig.~\ref{Fig.EnergyLS}
and investigate the change of $E0$ transition matrix element.
We introduce initial and final states.
The initial state, $\varPsi_i$, represents the first state 
of Fig.~\ref{Fig.EnergyLS}
and has compact tetrahedral structure.
The final state, $\varPsi_f$, represents the second state 
of Fig.~\ref{Fig.EnergyLS}
and has $^{12}\mathrm{C}+\alpha$ structure.
We define the initial state as $\varPsi_i\equiv\varPsi(R=0.1\,\mathrm{fm}, h=\sqrt{2/3}\times{}0.1\,\mathrm{fm}, \varLambda=0)$ and the final state as $\varPsi_f\equiv\varPsi(R=0.1\,\mathrm{fm}, h=\sqrt{2/3}\times{}5.0\,\mathrm{fm}, \varLambda=\varLambda_f)$.
Here $\varLambda_f$ is a control parameter, which changes the $^{12}$C cluster part
from three $\alpha$ to the subclosure of $p_{3/2}$ orbits.
Indeed, $\varPsi_f$ with $\varLambda_f=1$ is the dominant basis state for the second state
of Fig.~\ref{Fig.EnergyLS}, 
when the strength of the spin-orbit force is $V_\mathrm{LS}=3000\,\mathrm{MeV}$.\par
The $E0$ transition matrix element between the initial state $\varPsi_i$ and the final state $\varPsi_f$ is shown in Fig.~\ref{Fig.MonoVSLambda} as a function of $\varLambda_f$.
At $\varLambda_f=0$, the calculated $E0$ transition matrix element is $20.1\,e\,\mathrm{fm^2}$.
The value is about six times compared with the experimental one \cite{Ajzenberg_NuclPhysA460_1,Tilley_NuclPhysA564_1}.
However, this $E0$ transition matrix element drastically decreases with increasing $\varLambda_f$, and 
the value becomes comparable to the experiments around $\varLambda_f=1$.\par
\begin{figure}[tb]
\centering
\includegraphics[width=.48\textwidth]{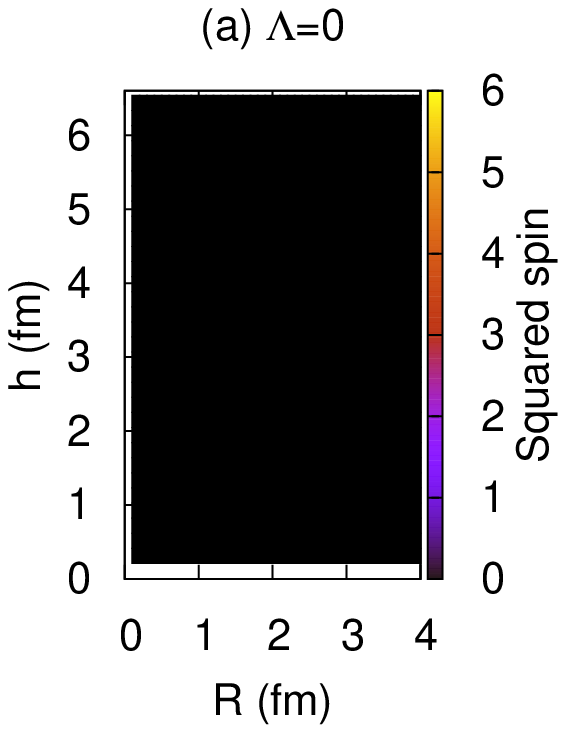}
\includegraphics[width=.48\textwidth]{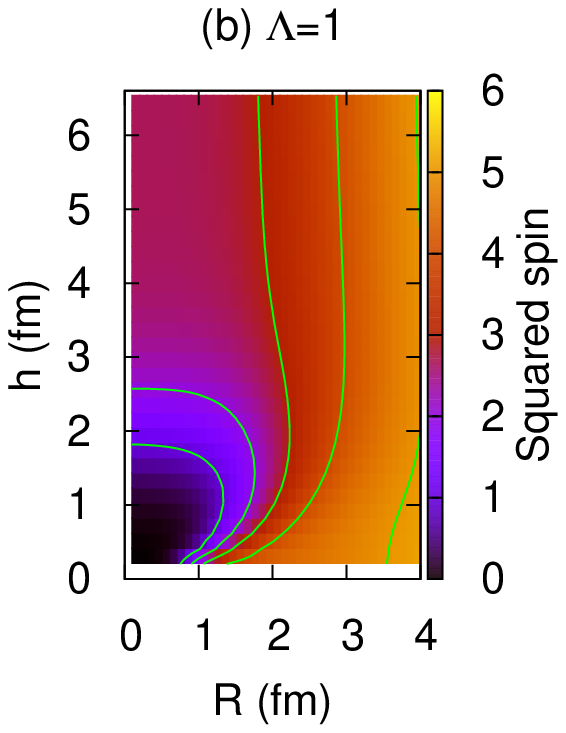}
\caption{(Color online) The expectation value of squared spin at (a) $\varLambda=0$ and (b) $\varLambda=1$ for the basis state $|\varPsi(R,h,\varLambda)\rangle$. }
\label{Fig.SquaredSpin}
\end{figure}

As a mechanism to explain this drastic decrease of the $E0$ transition matrix element
to the second $0^+$ state, 
we take notice on the change of the intrinsic spin structure in the final state.
The initial state is compact four $\alpha$ state, which agrees with the closed $p$ shell configuration, and 
the state has intrinsic spin equal to zero independent of the $\varLambda$ value. However the final state is $^{12}\mathrm{C}+\alpha$ cluster state, and when $\varLambda_f>0$, quasi $\alpha$ clusters can have finite expectation values of the intrinsic spin unlike $\alpha$ clusters.
Since the $E0$ transition operator does not contain the spin part, it cannot
connect two states with different intrinsic spin structures.\par
The intrinsic spin operator $\hat{\bvec{S}}$ is defined as
\begin{align}
\hat{\bvec{S}}=\sum_{i=1}^{16}\hat{\bvec{s}}_i,
\end{align}
where $\hat{\bvec{s}}_i$ is the spin operator for the $i$th nucleon.
Figure~\ref{Fig.SquaredSpin} shows the expectation value of the square of this operator as functions of $R$ and $h$.
Figure~\ref{Fig.SquaredSpin} (a) is the case of $\varLambda=0$, and the values are exactly zero independent on $R$ and $h$ values, because each $\alpha$ cluster has spin zero \cite{BrinkProceedings}.
On the contrary,  Fig.~\ref{Fig.SquaredSpin} (b) shows the case of $\varLambda=1$, and 
the values become finite except for the region of very small $R$ and $h$ values.
Although we do not change the spin part of the wave function,
the total system has finite intrinsic spin owing to the parameter $\varLambda$ 
given to the spatial part.
The expectation value of squared intrinsic spin for the final state with
$\varLambda_f=1$, which is the dominant basis state for the second state 
calculated with $V_\mathrm{LS}=3000\,\mathrm{MeV}$, is $2.6$.
Since the value for the ground state is almost zero, the two states have quite different spin structures.
The $E0$ transition operator does not act on the spin part, thus the $E0$ transition between two states 
which have different spin values is suppressed.
In Fig.~\ref{Fig.SquaredSpin} (b), the expectation value of squared spin numerically  converge to $2.66...
\approx{}8/3$ at the limit of $R\to0$ when $h$ value is large enough.

\subsubsection{Intrinsic spin of the full solutions}

\begin{table}[tb]
\centering
\caption{The expectation value of the square of the intrinsic spin operator for the $0^+$ states of $^{16}$O
obtained in Fig.~\ref{Fig.EnergyLS}.
Here $V_\mathrm{LS}$ stands for the strength of the spin-orbit force in the Hamiltonian.
}
\begin{tabular}{lrrrrrrr}
\hline\hline
$V_\mathrm{LS}\,\mathrm{(MeV)}$ & $0$ & $500$ & $1000$ & $1500$ & $2000$ & $2500$ & $3000$ \\ \hline
$0_{1}^+$ & $0.0$ & $0.0$ & $0.0$ & $0.1$ & $0.1$ & $0.2$ & $0.3$ \\
$0_{2}^+$ & $0.0$ & $0.0$ & $0.1$ & $0.1$ & $0.4$ & $1.2$ & $1.8$ \\
$0_{3}^+$ & $0.0$ & $0.1$ & $0.3$ & $0.7$ & $0.9$ & $0.6$ & $1.9$ \\
$0_{4}^+$ & $0.0$ & $0.0$ & $0.0$ & $0.0$ & $1.8$ & $1.9$ & $0.7$ \\ \hline\hline
\label{ssq}
\end{tabular}
\end{table}

Next, we discuss the intrinsic spin of the full solution.
The expectation value of the square of the intrinsic spin operator for the full solutions 
obtained in Fig.~\ref{Fig.EnergyLS},
are shown
in Table~\ref{ssq}.
At $V_\mathrm{LS} = 0$ MeV, all the $0^+$ states listed here have the value close to zero.
With increasing $V_\mathrm{LS}$, the basis states with
finite $\varLambda$ start contributing to each state, and the intrinsic spin increases
in all the states listed here. However the increase is much smaller in the ground state
due to the closed shell structure of the $p$ shell;
the state has the value of 0.3 at $V_\mathrm{LS} =3000$ MeV.
On the other hand,
the value for the second $0^+$ state is 1.8, where the subclosure configuration of
the $p_{3/2}$ shell for the $^{12}$C cluster part is important. 
The intrinsic spin structure
of these two states are completely different. 
This is the reason for the 
suppression of the $E0$ transition probability between these states.
Although 
the $E0$ operator itself does not have the spin dependence, the matrix element
is sensitive to the difference of intrinsic spin structures of the two states.

\subsection{Orientation of $^{12}$C}
In our model, we considered only triangular pyramid structures, \textit{i.e.} we did not consider the effect of rotation of three $\alpha$ clusters forming $^{12}\mathrm{C}$ with respect to the last $\alpha$ cluster.
The definition of the coordinate system in this work may lead to the overestimation of the $E0$ transition strength. 
Taking into account other orientations of $^{12}\mathrm{C}$ as in the previous study \cite{En'yo_PhysRevC89_024302} is expected to help in better reproduction of the experimental value.\par
For instance, 
in Fig.~\ref{Fig.MonopoleLS},
the  $E0$ transition matrix elements from the first 
to the second $0^+$ states  is around 8.3 $e$~fm$^2$ 
at $V_\mathrm{LS}$ = 0 MeV. This value is reduced when other orientations of $^{12}$C
are introduced. If we prepare $\varLambda = 0$ (four $\alpha$)
wave functions with other orientations of $^{12}$C with respect to the forth $\alpha$
and diagonalize the Hamiltonian, the value decreases to 6.9  $e$~fm$^2$.
In our analyses we needed rather large value of 
$V_\mathrm{LS} \sim $ 2500 MeV to reproduce the experimental 
$E0$ value from the ground to the second $0^+$ state;
however this result indicates that we could reproduce it with a bit smaller 
$V_\mathrm{LS}$ value when this orientation effect is taken into account.

\section{conclusion}
\label{conclusion}
In this study, for  $^{16}\mathrm{O}$, the $0^+$ energy levels and the $E0$ transition matrix elements from the ground state have been investigated in the framework of AQCM.
The $E0$ transition strength has been known as a quantity which characterizes the cluster structure of low-lying excited states, and here we focused on the dependence on the strength of the spin-orbit force, $V_\mathrm{LS}$.\par
The ground state is compact four $\alpha$ state and almost independent of $V_\mathrm{LS}$. 
On the contrary, as pointed out by many previous works, cluster structure is important in the $0_2^+$ state, and this is obtained also in our model.
In addition, in the present study we discussed the change of the wave function of 
the $^{12}\mathrm{C}$ cluster part. 
With increasing $V_\mathrm{LS}$, the level repulsion occurs and the $^{12}\mathrm{C}$ cluster part change from three $\alpha$'s, which are not affected by the spin-orbit force, to the $p_{3/2}$ subclosure of the $jj$-coupling shell model, and the excitation energy of the $0^+_2$ state drastically decreases.\par
The $E0$ transition matrix elements from the ground state to the excited states are strongly dependent on the level repulsions.
For $0\le{}V_\mathrm{LS}\le1500\,\mathrm{MeV}$, the $E0$ transition matrix element from the ground state to the second $0^+$ state is around 8 $e$~fm$^2$, which is much larger than the observed one ($3.55\pm{}0.21\,e~\mathrm{fm}^2$).
In the region of $V_\mathrm{LS}\ge1500\,\mathrm{MeV}$, it starts decreasing and
becomes comparable to the experimental one slightly above  $V_\mathrm{LS} = 2000\,\mathrm{MeV}$.
On the other hand, the value from the ground state to the third $0^+$ state 
is around 2 $e$~fm$^2$ at $V_\mathrm{LS} = 0$ MeV 
and this is too small compared with the experimental value of
$4.03\pm0.09\,e~\mathrm{fm}^2$.
The value increases and becomes comparable to the experimental one
around $V_\mathrm{LS} = 2000$ MeV.
The decrease of the transition matrix element to the second $0^+$ 
and the increase to the third $0^+$ state
is due to the level repulsion between the second and third states.
The wave functions of these states are interchanged. 
Experimentally the transition matrix element to the third $0^+$ state is slightly larger
than one for the second $0^+$ state,
and this is realized with the $V_\mathrm{LS}$ value slightly above 2000 MeV.
In this model, only triangular pyramid structure of $^{16}$O has been considered; however
taking into account other orientations of $^{12}\mathrm{C}$ part with respect to $\alpha$
is expected to help in further reproduction of the experimental transition matrix elements.
\par
The expectation value of the square of the intrinsic spin was also analyzed.
At $V_\mathrm{LS} = 0$ MeV, the ground and second $0^+$ states have the value close to zero.
With increasing $V_\mathrm{LS}$, the basis states with
finite $\varLambda$ start contributing to each state, and the intrinsic spin increases;
however the increase is much smaller in the ground state
due to the closed shell structure of the $p$ shell
(the value is 0.3 at $V_\mathrm{LS} =3000$ MeV).
On the other hand, at $V_\mathrm{LS} =3000$ MeV,
the value for the second $0^+$ state is 1.8, where the subclosure configuration of
the $p_{3/2}$ shell for the $^{12}$C cluster part is important. 
The intrinsic spin structure
of these two states are completely different. 
This is the reason for the 
suppression of the $E0$ transition probability between these states.
Although 
the $E0$ operator itself does not have the spin dependence, the matrix element
is sensitive to the difference of intrinsic spin structures of the two states.\par
In the traditional microscopic $\alpha$ cluster models,
there has been a long standing problem that the calculated excitation energy 
of $0^+_2$  is higher than the experiments by more than $10\,\mathrm{MeV}$,
since the spin-orbit force is missing.
Now this is considerably improved by introducing the dissolution of $\alpha$ clusters.
However, taking into account other effect, such as three-body force effect \cite{Itagaki_ProgTheorPhys94_1019}, 
would be promising in order to fully solve the problem.\par
In this study we found an important correlation between the strength of the spin-orbit force and $E0$ transition to low-lying excited states.
The $E0$ transition is sensitive to the persistence of $N\alpha$ correlation in the excited states and can be its measure.
Similar investigations for other light nuclei are on going.

\begin{acknowledgments}
The authors would like to thank T. Ichikawa for the discussions.

\end{acknowledgments}

\end{document}